\documentstyle[12pt,aasms4]{article}

\begin{document}
\title{A TEST OF THE PARTICLE PARADIGM IN N--BODY SIMULATIONS}
\author{Bryce Kuhlman, Adrian L. Melott, and Sergei F. Shandarin}
\affil{Department of Physics and Astronomy, University of Kansas, Lawrence, KS
66045}
\affil{kuhlman@, melott@, sergei@kusmos.phsx.ukans.edu}

\begin{abstract}
We present results of tests of the evolution of small ``fluid elements" in
cosmological N--body simulations, to examine the validity of their treatment as
particles. We find that even very small elements typically collapse along one
axis while expanding along another, often to twice or more their initial
comoving
diameter. This represents a possible problem for high--resolution uses of such
simulations.

Subject headings:\ \ cosmology:miscellaneous--
gravitation--hydrodynamics--methods:numerical--dark matter
\end{abstract}
One of the bases of validity for cosmological N--body simulations concerns the
``particle" approximation. Such simulations are often taken to represent
cosmological gravitational clustering of
nonbaryonic dark matter.
In the
simulations, a particle acts as a substitute or ``phase space marker" for a
large
number of dark matter particles, stars, or possibly galaxies. With progress
in computing, the mass of such particles has dropped steadily, but still
represents a sizeable fraction of a galaxy mass, far more than that of an
elementary particle or even a star. One of the problems related to such
massive
particles is the phenomenon of gravitational collisions in N--body codes
recently emphasized by Suisalu \& Saar 1996
(see also the references therein).
It of course does not occur between individual dark
matter particles.

We distinguish two aspects of the discretization of cosmological simulations.
Particles
in the N-body simulations are supposed to move as an almost perfectly
continuous medium of dark matter on the one hand and
be the sources of gravitational force on the other.
In certain types of  numerical codes
(the so called high resolution schemes) the resolution of calculating the
force is substantially higher than that of the mass density.
This means that the force between the particles is approximated
by the Newton law for point--like masses. In reality the force must be
calculated as the interaction between clouds in the Eulerian space
which are maps of the initial clouds in the Lagrangian space.
The sizes of the Eulerian clouds are comparable to the
separation between the neighboring clouds.

It is then appropriate to ask whether the particle approach is valid.
Validity can
have many different meanings; we choose to define it as representing the
solution that would be reached in the limit of a very much larger set of
computational particles, after that solution is smoothed on some scale. Results
are often presented probing scales much smaller than the initial
(Lagrangian) mean interparticle
separation. This is questionable especially when dealing with nonbaryonic dark
matter (such as axions or massive neutrinos).

One issue that affects this question is compactness. Can a particle stand in
for an ensemble of smaller particles? One condition that affects this is
whether that set of particles move together, such that their replacement by a
single mass point is reasonable. It is often argued that at the nonlinear stage
of gravitational clustering most of the mass is in dense clumps. Therefore
particles are packed closely and fluid particles have smaller volumes than
initially. The Liouville Theorem combined with gravity assures that in clumps
the
physical volumes are smaller, but they may or may not be compact, as the phase
space is mixed. Furthermore, the physical volume may not become uniformly
smaller as the simulation progresses.
Although these assumptions look reasonable they have not been tested. Moreover,
a
careful study of the first collapse suggests that it is anisotropic in most
cases (Shandarin et al. 1995).

We examine this question by examining histories in a simulation. We study the
behavior of small ``clouds" in the course of evolution. We study two type of
clouds: the smallest possible volumes in the simulations made of $2^3=8$
neighboring
particles and larger clouds made of $5^3=125$ particles. Initially the clouds
have a cubic shape. Small clouds probe the smallest scale of the simulation but
do not feel inhomogeneities on even smaller scales that are always assumed in
realistic cosmological models but are beyond reach of simulations. Large
clouds are sensitive to the small--scale perturbations but evolve less than the
small ones. This example
is taken from a pure power--law spectrum $P(k)\propto k^{-1}$ set of initial
conditions. One advantage of power--law initial conditions is that they permit
a self--consistency test by examining results on different scales, to help
assure that resolution and boundary conditions are not contaminating the
results. These are $128^3$ Particle--Mesh simulations, described more fully in
Melott and Shandarin (1993), hereafter MS.

We chose 100 particles at random and examined the evolution of their
neighborhoods. One can imagine these as standing in for a single particle
in a simulation with less mass resolution.
We rejected any particles which did not lie in regions of
density contrast $\delta >10$ on the mesh scale at the last stage since we
wanted to
include only particles in fairly dense environment which is typical
for forming objects. Figures 1a--1c show the evolution of
three $2^3$ sets. Although not extermely high, the threshold $\delta >10$
selects isolated
(although not all of them necessarily gravitationally bound)
clumps at all stages;
the percolation threshold is much lower
(Yess \& Shandarin 1996).

In Figure 1(a), the volume element simply collapses. In this case, we see the
maximum validity of the particle approximation. Figure 1(b) shows a more
questionable situation. The element ends in a fairly compact state, but passes
through an intermediate pancake--like configuration
during the shell crossing stage.
Such configurations are a
generic result of gravitational instability, as predicted by the Zel'dovich
(1970) approximation but accurate for a far wider array of initial conditions
(Melott and Shandarin 1990;
Little et al. 1991; Coles, et al 1993, Bond et al. 1996).
Although the element becomes compact, for a time at least the quadrapole
moment of its potential is incorrect if  replaced by a single particle, which
can certainly affect torque on neighboring masses. In Figure 1(c) the fluid
element simply expands. It would be very poorly represented by a single mass
point. It is worth stressing that the PM code used in this study does not
particularly suffer from excessive scattering.

A $2^3$ particle array is close to the resolution limit of a PM code, so one
might worry that this strongly affects the results. In Figure 2a--2c we show
similar histories for three $5^3$ elements, with similar characteristics
(although of course they collapse later).

It is important to quantify these histories in some way, and ask how typical
they are. We do this by using eigenvalues and eigenvectors of the tensor

\begin{equation}
M_{ij}=\sum\limits^N_{n=1} r_i^{(n)}r_j^{(n)}
\end{equation}
\noindent where $i$ and $j$ run through coordinate indices 1 to 3, coordinates
measured relative to the center of mass, and $n$ is
the number of particles in the object (8 or 125). The eigenvectors of $M$ are
of course principal
axes. The square
roots of the eigenvalues of $M$ are proportional to the axis lengths if the
particles are scattered uniformly in an ellipsoid. Figures 1 and 2 are in fact
viewed along eigenvectors of $M$ for the final configuration, showing it from
perspectives which maximize flattening.

We examine the evolution of the eigenvalues of $M$ from the initial conditions
to a deeply nonlinear state. Our most nonlinear stage corresponds to
fluctuations of order unity when the evolved density field is smoothed with a
tophat of radius $L/5$, where $L$ is the diameter of the (128$^3$ cell)
simulation. We
took a random sample of 100 particles from overdense regions and followed the
evolution of their neighborhoods. Figure 3 shows the evolution of distribution
of the largest
(c) and smallest (a) eigenvalues up to this
stage (in expansion factors of two) as well as for the slightly deformed $2^3$
lattice of initial conditions. We have taken square roots and normalized so
that the number may be considered an axis length in cell units for ellipsoids.

The shortest axis undergoes collapse for almost all such elements. The collapse
of the shortest axis proceeds fairly fast from the initial stage until the
scale
of nonlinearity reaches $k^{-1}_{n\ell}\approx 2.5$ (here 1 is the initial
particle separation on the mesh).
They seem to reach a stable comoving size by this stage, which we can
crudely label virialization. The median element has reduced its short axis by
about
four times.

Meanwhile the longest axis typically expands. (All the following statements
are in comoving coordinates.)
For approximately 90\% of the
elements this axis has grown at the moment of collapse. For about one--third of
them it has more than doubled. The long axis also shows a jump in its behavior
(more small values) just at the moment we identify as the onset of
nonlinearity. This probably corresponds to some merging. But still more than
two--thirds of the elements are expanded along this axis, and would be
questionable candidates for replacement by single particles.

In Figure 4, we show the evolution of axes for our $5^3$ elements. With
adjustments to the epoch of nonlinearity for the larger volume element, they
confirm the trends found in the smaller volume. Here the distribution function
stops evolving (or at least slows down evolving) at $k^{-1}_{n\ell}\simeq 5$.
Recalling that the plots of the distribution function are given at discrete
instants of time, one can say that there is rough (but  not perfect) scaling.
This time the median element has reduced its short axis by about three times.
This difference with $2^3$ may be nothing more than a resolution effect on the
smaller volumes, whose short axes collapse below our force resolution limit.
More importantly, at the collapse
epoch (about the fourth evolved stage) the long axis has grown for $\sim$ 90\%
of volume elements, and about half have more than doubled. We conclude that
shear is
important in dense regions.

We do not wish to imply that N--body simulations are necessarily invalid. It is
possible that an ensemble of computer particles may (or may not) act like an
ensemble of dark matter particles.
Some
of the effect we are seeing is orbit mixing inside extended bound objects. On
the other hand, even for the $2^3$ elements some substantial fraction of axes
are longer than those of any cluster, and larger than the mass autocorrelation
length. Only a small fraction of fluid particles that finally found
themselves in a fairly dense environment $(\delta >10)$ has collapsed along
all three directions by a factor of two. At least 50\% of such particles have
the longest size as large as the initial size. These results are in good
qualitative agreement with figures presented by Gnedin (1995).

Simulating the dark matter distribution (e.g, axions, massive neutrinos), we
deal with an almost perfect continuous medium. Since scattering effects between
individual dark matter particles are small, two infinitesimally close fluid
particles
should remain infinitestimally close, although the separation may increase.
The most conservative approach would be to restrict one's attention to volumes
large enough to encompass nearly all long axes. In practical terms, this is
several times the mean interparticle separation  (with dependence
on the type of spectrum and degree of nonlinearity). We have not definitively
shown that there are fatal errors in the N--body approach.
However, we present these
results to be considered further as we move into a new era of high
resolution simulations, with N--body dark matter providing the background for
hydrodynamics work on small scales. A possible approach which is free
from this problem is a Poisson solver in a quasi--Lagrangian
space (Gnedin, 1995). However, in its current form it is limited to
times prior to the first crossing of particle orbits, which is very
early.   Splinter (1996) has developed a nested--grid code which
puts a higher particle density in regions of interest, greatly reducing
the discreteness problem in those regions.

\acknowledgements
{We thank Jenny Pauls for the use of her software, NASA grant NAGW--3832 and
NSF
support of Research Experiences for Undergraduates, and F. Bouchet for
discussion and valuable comments.  Simulations were performed at the
National Center for Supercomputing Applications, Urbana, IL.}

\newpage
\figcaption[] { Three views of a collapsing $2^3$ particle volume element
along the three principal axes of its second moment.
Views c--a are along the long through short axes, respectively.  The top
row represents the initial conditions.
Steps that follow are spaced by an expansion factor of two
in the background cosmology.  (a) This element collapses nicely to a compact
object. (b) This element passes through a pancake phase before reaching
compactness. (c) This element is scattered, growing in size even in comoving
coordinates. }
\figcaption[] {Same as Figure 1, except the views are of $5^3$ particle
element, and the scale is expanded appropriately.}
\figcaption[] { The evolution of the small (a) and large (c) principal
axis lengths for a subsample of 100 particle Lagrangian $2^3$ neighborhoods as
described in the text. The evolution is shown from initial conditions (dotted
line) and evolved stages in the order shortdash, longdash, dot--shortdash,
dot--longdash, shortdash--longdash, solid line, corresponding to the stages
used in MS for $n=-1$. All of the evolved stages are separated by an expansion
factor of two.}
\figcaption[] {Same as Fig. 3, except for a $5^3$ particle neighborhood.}

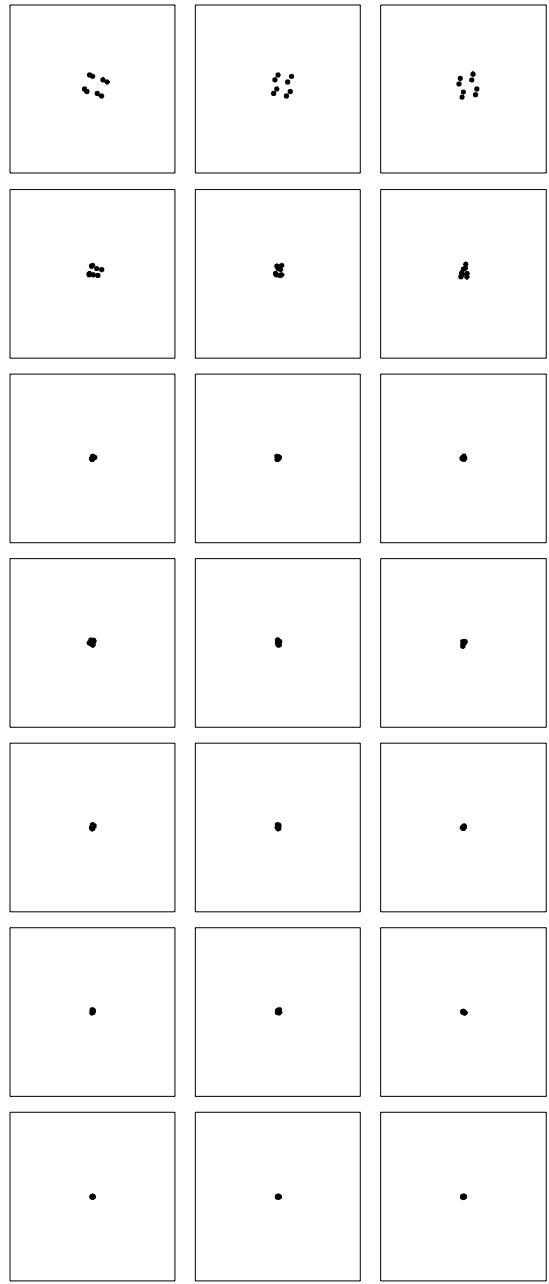
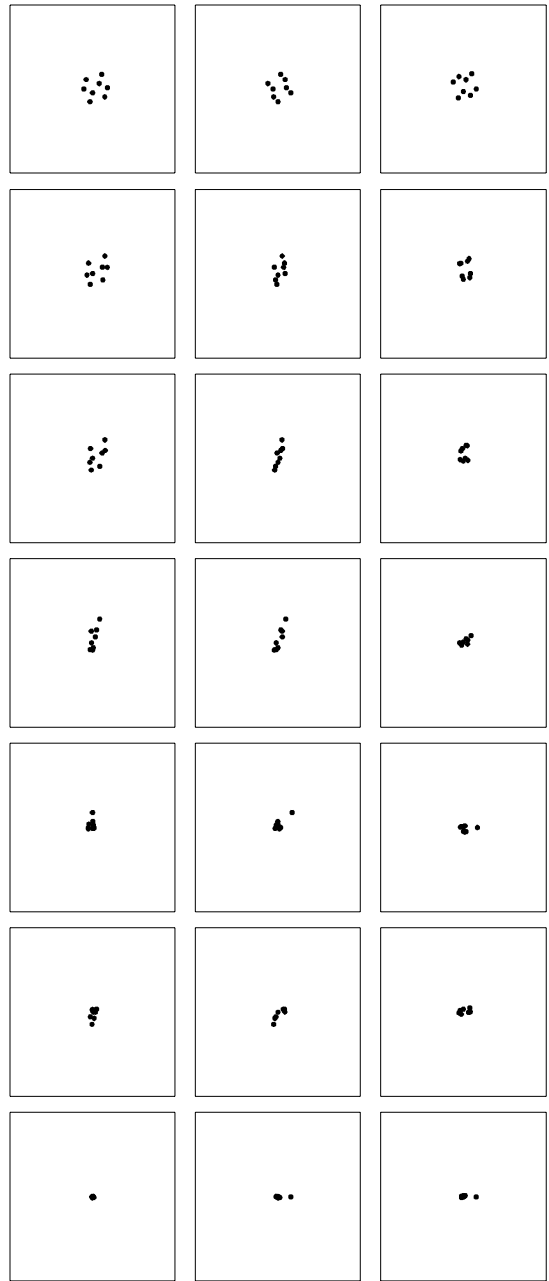
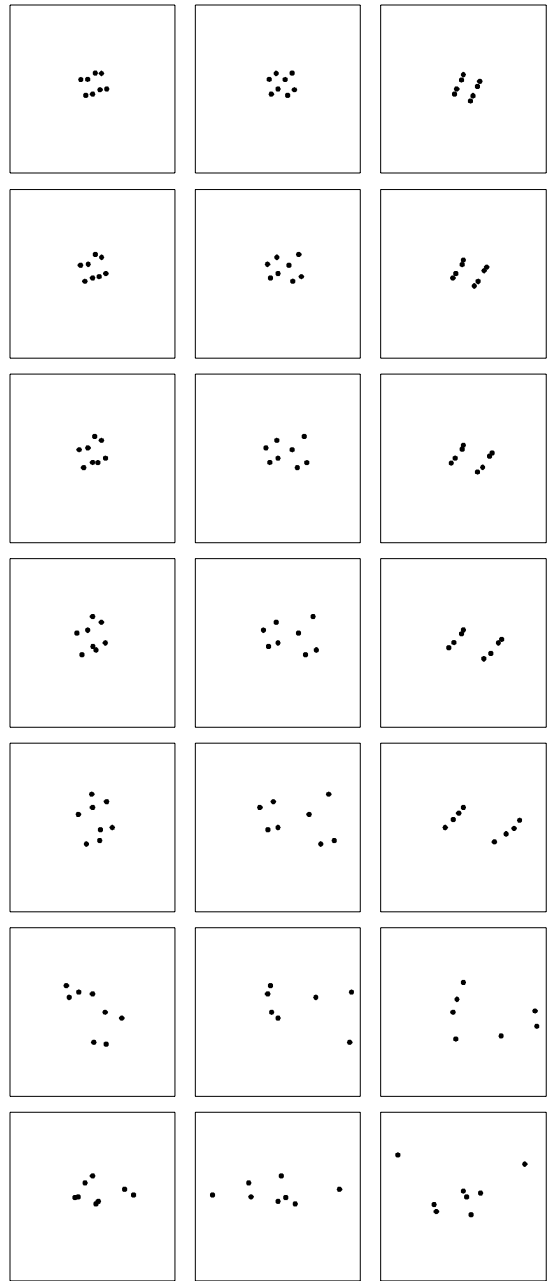
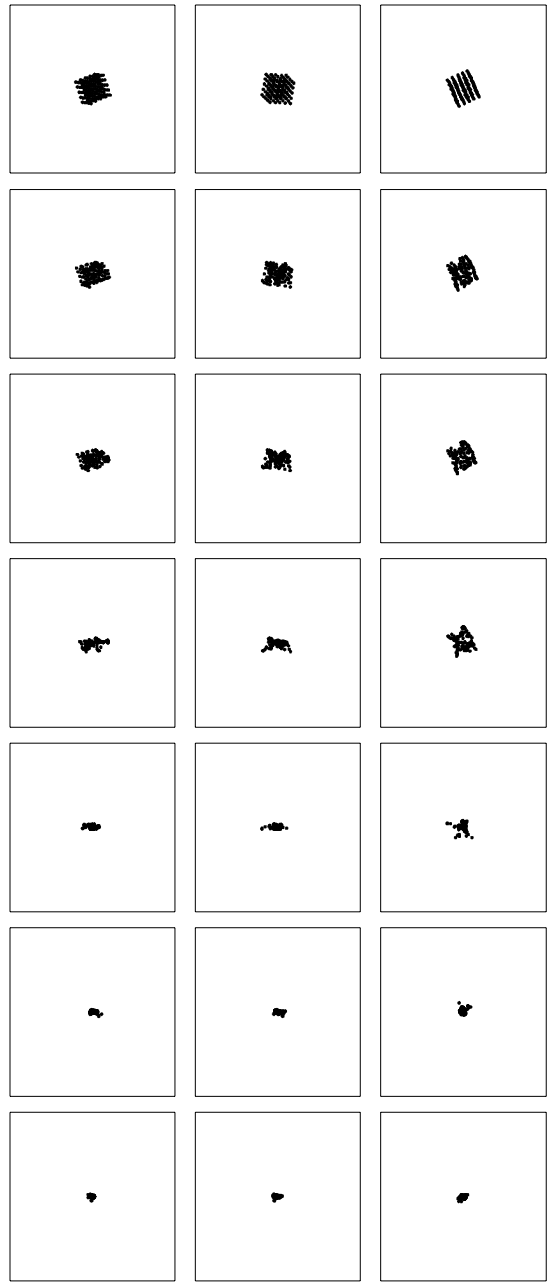
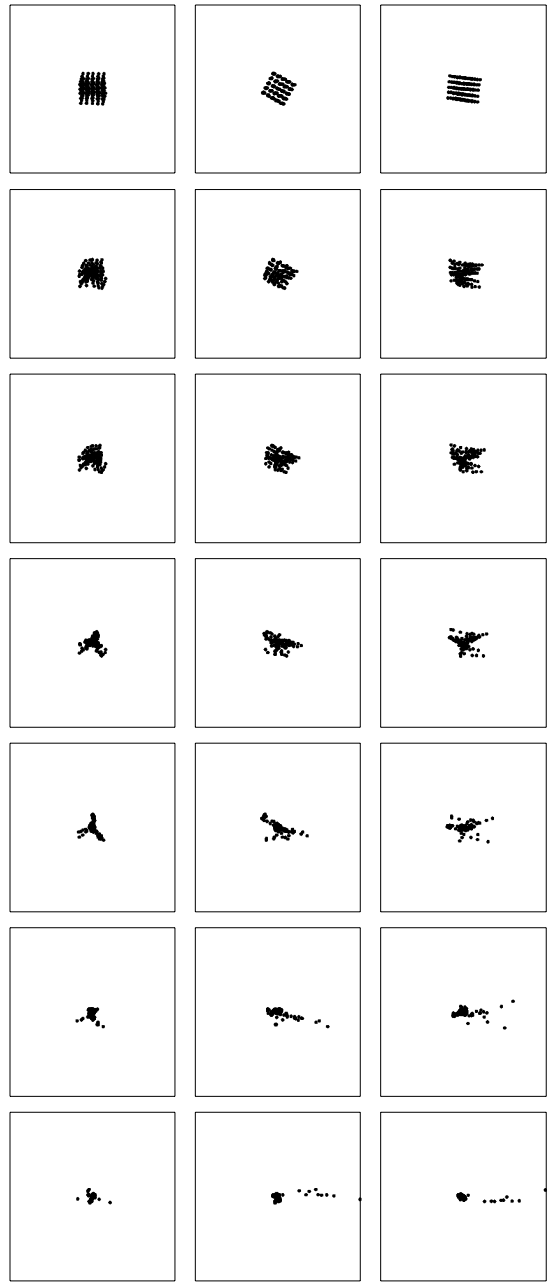
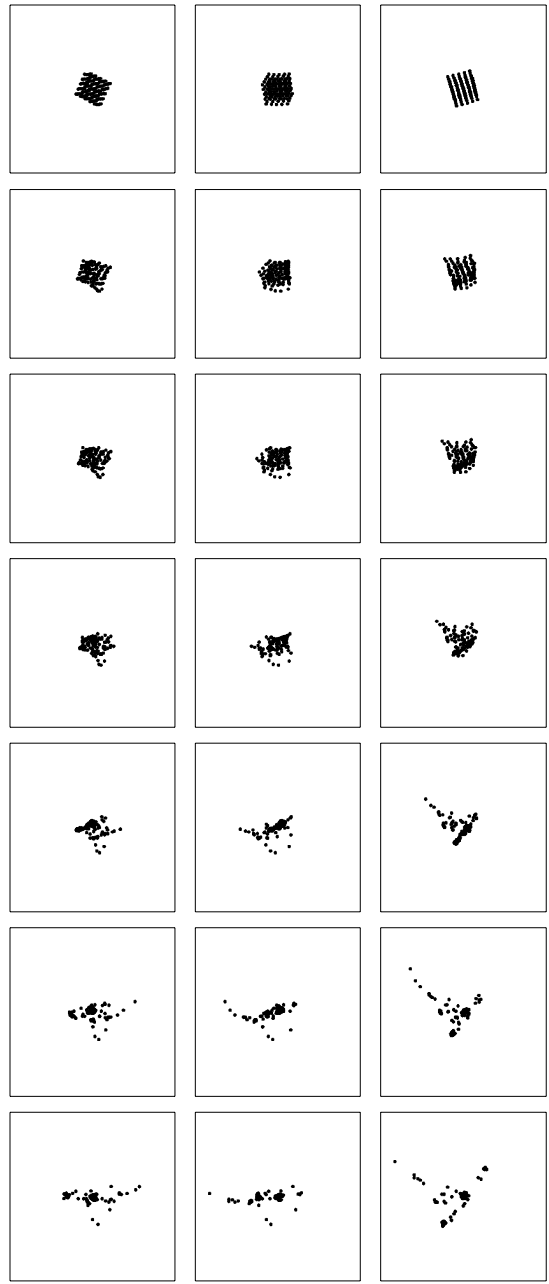
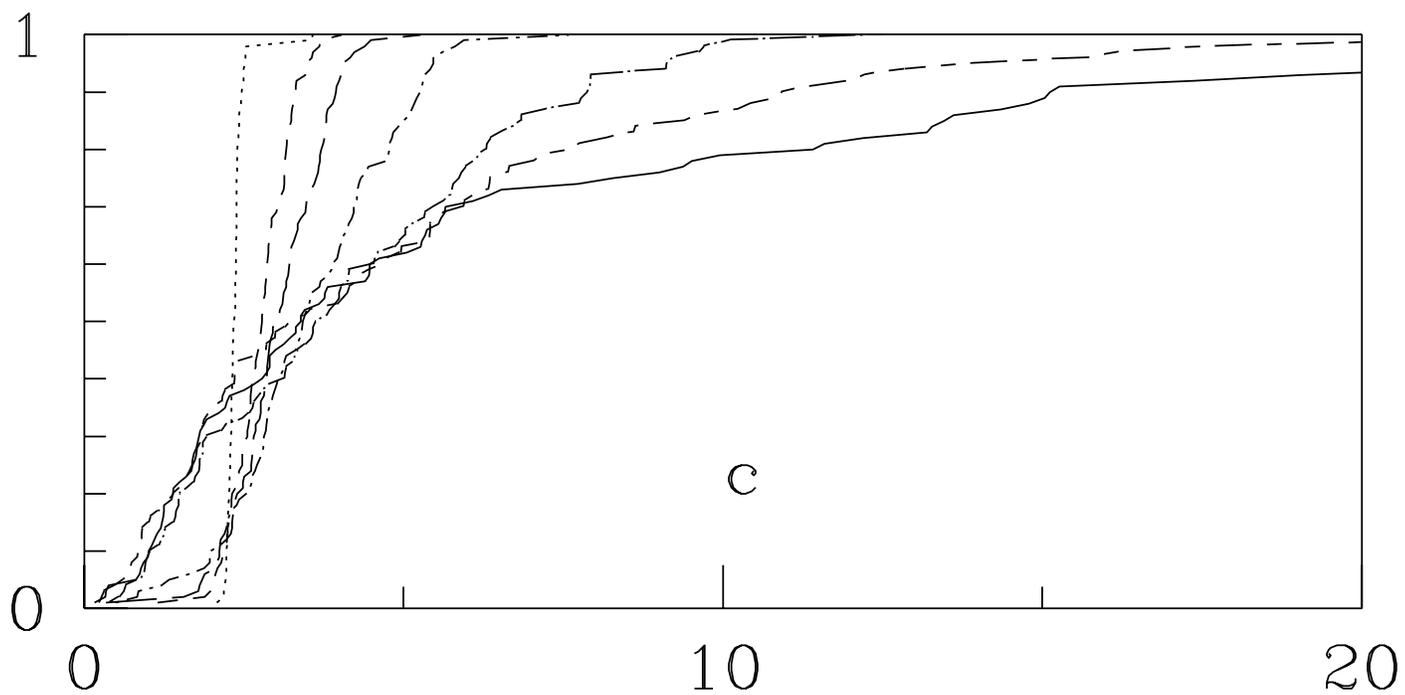
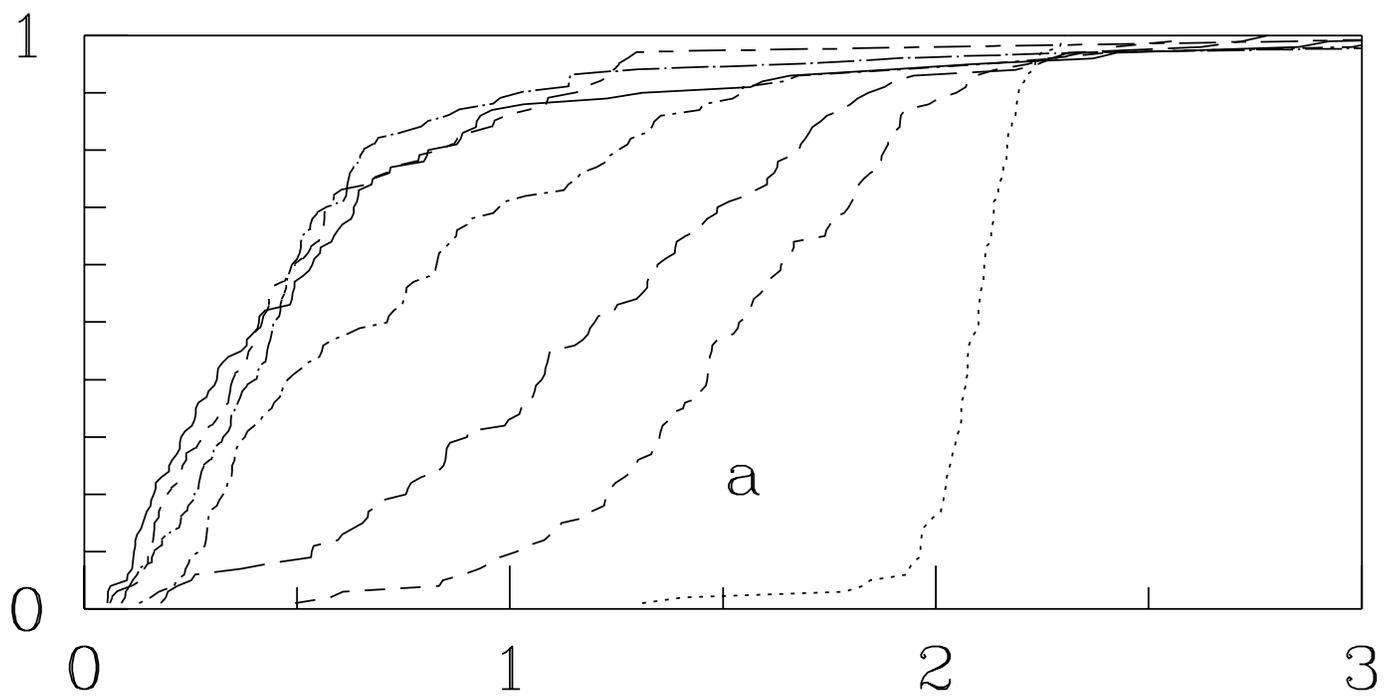
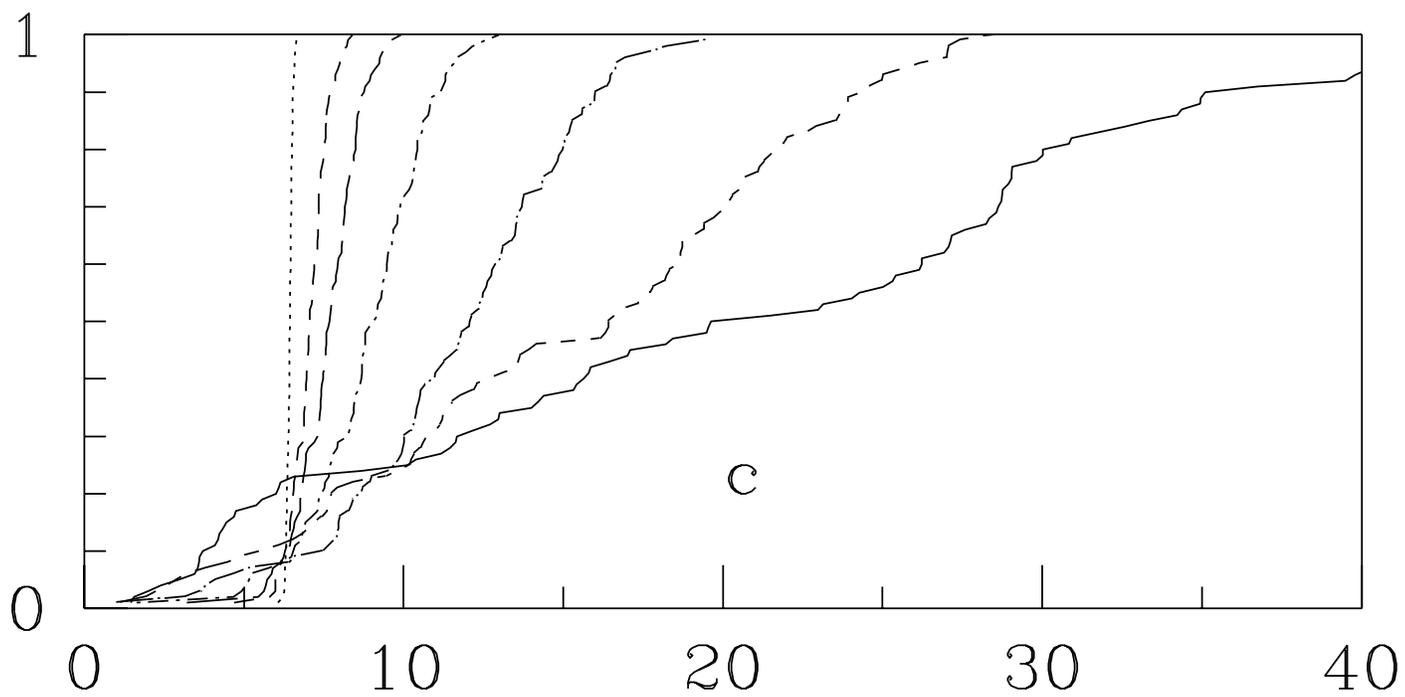
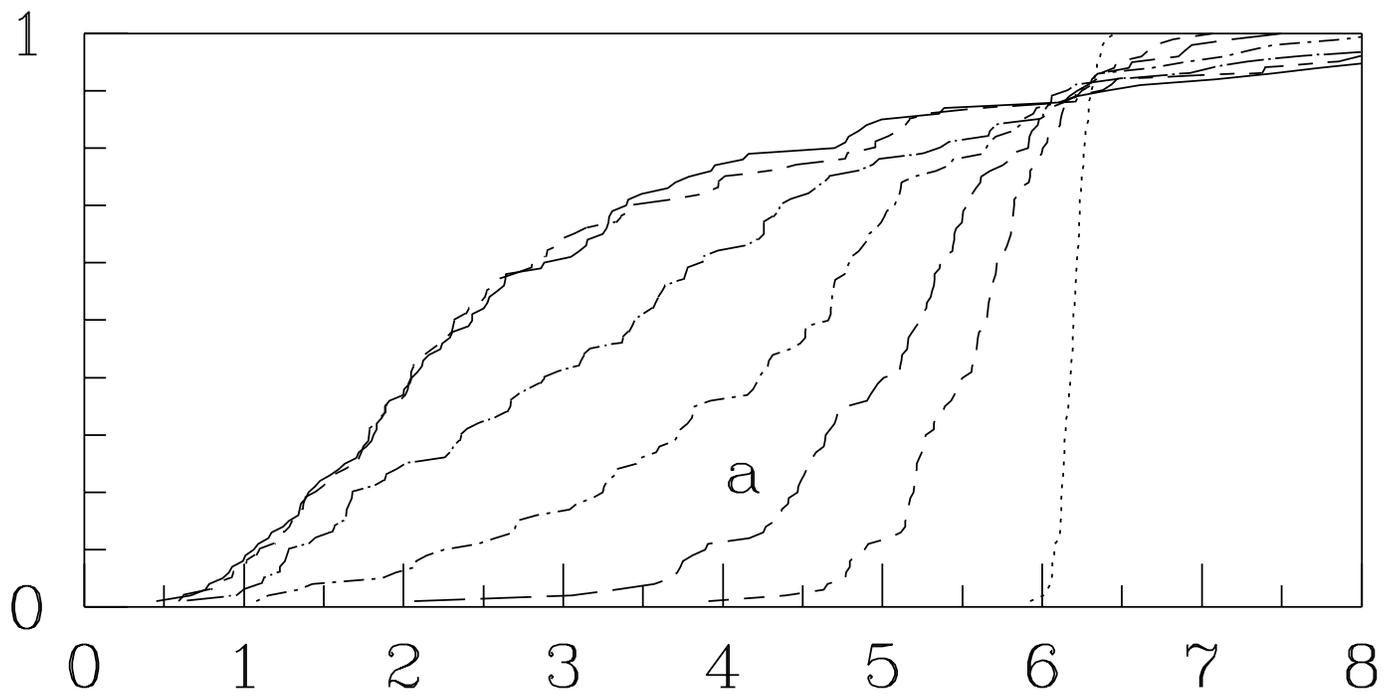
\end{document}